\documentclass[12pt]{article}
\usepackage{amsmath}
\usepackage{amssymb}
\usepackage{latexsym}

\begin{document}

\baselineskip 6mm
\renewcommand{\thefootnote}{\fnsymbol{footnote}}

\newcommand{\nc}{\newcommand}
\newcommand{\rnc}{\renewcommand}

%\headheight=0truein
%\headsep=0truein
%\topmargin=0truein
%\oddsidemargin=0truein
%\evensidemargin=0truein
%\textheight=9truein
%\textwidth=6.5truein

\rnc{\baselinestretch}{1.24}    % 1.5 spacing btwn text lines
\setlength{\jot}{6pt}       % spacing btwn the rows of an eqnarray
\rnc{\arraystretch}{1.24}   % spacing btwn the rows of a non-eqn array

%%%%%%%%%%%%%%%%%%%%%% Equation Numbering %%%%%%%%%%%%%%%%%%%%%%%
%\makeatletter
%\rnc{\theequation}{\thesection.\arabic{equation}}
%\@addtoreset{equation}{section}
%\makeatother

%%%%%%%%%%%%%%%%%%%%%%%%%%%%%%%%%%%%%%%%%%%%%%%%%%%%%%%%%%%%%%%%%
%                                                               %
%                NEW COMMANDS AND MACROS                        %
%                                                               %
%%%%%%%%%%%%%%%%%%%%%%%%%%%%%%%%%%%%%%%%%%%%%%%%%%%%%%%%%%%%%%%%%

%%%%% Simplify some frequently used LaTeX commands %%%%%

\nc{\be}{\begin{equation}}

\nc{\ee}{\end{equation}}

\nc{\bea}{\begin{eqnarray}}

\nc{\eea}{\end{eqnarray}}

\nc{\ben}{\begin{eqnarray*}}

\nc{\een}{\end{eqnarray*}}

\nc{\xx}{\nonumber\\}

\nc{\ct}{\cite}

\nc{\la}{\label}

\nc{\eq}[1]{(\ref{#1})}

\nc{\newcaption}[1]{\centerline{\parbox{6in}{\caption{#1}}}}

\nc{\fig}[3]{

\begin{figure}
\centerline{\epsfxsize=#1\epsfbox{#2.eps}}
\newcaption{#3. \label{#2}}
\end{figure}
}

%%% Double line letters %%%

\def\IR{{\hbox{{\rm I}\kern-.2em\hbox{\rm R}}}}
\def\IB{{\hbox{{\rm I}\kern-.2em\hbox{\rm B}}}}
\def\IN{{\hbox{{\rm I}\kern-.2em\hbox{\rm N}}}}
\def\IC{\,\,{\hbox{{\rm I}\kern-.59em\hbox{\bf C}}}}
\def\IZ{{\hbox{{\rm Z}\kern-.4em\hbox{\rm Z}}}}
\def\IP{{\hbox{{\rm I}\kern-.2em\hbox{\rm P}}}}
\def\IH{{\hbox{{\rm I}\kern-.4em\hbox{\rm H}}}}
\def\ID{{\hbox{{\rm I}\kern-.2em\hbox{\rm D}}}}

%%%%% Roman pont in math

\def\Tr{{\rm Tr}\,}
\def\det{{\rm det}}

%%%%% Special Letters

\def\vare{\varepsilon}
\def\barz{\bar{z}}
\def\barw{\bar{w}}

%\begin{titlepage}
%---------------- preprint number ---------------
\hfill\parbox{5cm}
{DAMTP-2004-149 \\ \\ hep-th/0412153}\\
\vspace{25mm}
\begin{center}
%------------------------ title ------------------------
{\Large {\bf Electrically charged dilatonic black rings}  }

\vspace{15mm}
%---------------- authors and addresses ----------------
Hari K. Kunduri \footnote{H.K.Kunduri@damtp.cam.ac.uk} and James Lucietti\footnote{J.Lucietti@damtp.cam.ac.uk}
\\[3mm]
{\it  DAMTP, Centre for Mathematical Sciences,\\
      University of Cambridge,
      Wilberforce Rd.,\\
      Cambridge CB3 0WA, UK\\}

\end{center}

\thispagestyle{empty}

\vskip2cm

%----------------------- abstract ----------------------

\centerline{\bf Abstract}
\vskip 4mm
\centerline{
\parbox[t]{15cm}{\small
\noindent In this note we present (electrically) charged dilatonic black ring solutions
of the Einstein-Maxwell-dilaton theory in five dimensions and we consider
their physical properties. These solutions are static and as in the
neutral case possess a conical singularity. We show how one may
remove the conical singularity by application of a Harrison
transformation, which physically corresponds to supporting the charged
ring with an electric field. Finally, we discuss the
slowly rotating case for arbitrary dilaton coupling. }}

\vspace{2cm}

%\end{titlepage}

%\section{Introduction}

\noindent One of the most exciting recent results in higher
dimensional gravity was Emparan and Reall's discovery of the  black
ring in~\cite{GEN,ROT}. Their solution describes the gravitational
field of an isolated source equipped with an event horizon of $S^1
\times S^2$ topology. Although the static black ring is plagued by a
conical singularity, the authors were subsequently able to show the
existence of a vacuum rotating black ring~\cite{ROT}. Since then,
more solutions have been found in various five dimensional gravity
theories. Elvang was able to apply a Hassan-Sen transformation to
the solution~\cite{ROT} to find a charged black ring in the bosonic
sector of the truncated heterotic string theory~\cite{SCR}.
Furthermore, a supersymmetric black ring in five dimensional minimal
supergravity was derived~\cite{SUSYRING} and then generalized to the
case of concentric rings in~\cite{CONRING}. Finally, these solutions
were placed into the context of string and M theory, corresponding
to D1-D5-P supertube configurations~\cite{SRING}.

Despite the number of interesting solutions, it is rather surprising
that an electrically charged rotating black ring in the well-known
Einstein-Maxwell (EM) theory remains to be found. In~\cite{IDA} a
static black ring carrying electric charge was given, however their
expression for the field strength is incorrect. In~\cite{MAGRING}
Emparan was able to derive ``dipole black rings'' in the
Einstein-Maxwell-dilaton (EMD) theory in five dimensions. Such
solutions are supported by magnetic potential (though they are
electrically charged by the Kalb-Ramond two form potential in the
dual theory).

In this paper we consider electrically charged black rings in the
EMD theory. Our motivation is twofold: firstly, such solutions are
interesting in their own right. Secondly, it has been shown that in
a particular degenerate limit (rotating) black rings turn into
(rotating) black holes. In particular the rotating black ring of
pure 5D gravity reduces in this limit to the Myers-Perry black
hole~\cite{MYERSPER} with one non-vanishing angular momentum. As yet
to the best of our knowledge there is no general Maxwell-charged
Myers-Perry solution found in the literature; such a solution would
be the higher dimensional counterpart of the Kerr-Newman solution in
four dimensions\footnote{It should be noted that in~\cite{pope}, a
rotating charged black hole solution of EM theory with a
Chern-Simons term can be found where the rotation parameters are
equal.}. In this note we begin this line of attack by presenting a
static charged dilatonic black ring valid for all values of the
dilaton coupling.

This paper is structured as follows. Firstly, we present our
solutions and show that they may be interpreted as static, charged,
dilatonic black rings. We discuss their physical properties and show
they do not possess a sensible extremal limit. As in the neutral
case, these static black rings suffer from the existence of a
conical singularity. Next, we consider the possibility of removing
the conical singularity and apply certain solution generating
techniques in order to find a new static electrically charged black
ring immersed in a background electric field. Finally, we conclude
with a discussion on the possibility of finding rotating (charged
electrically under the
Maxwell field) dilatonic black rings. \\ \\

%\section{Charged dilatonic black rings}

We consider the EMD system, in $4+1$
dimensions defined by the following action: \be \label{action} S = \frac{1}{16\pi }\int d^5x
\sqrt{-g}( R - 2g^{\mu\nu}\partial_{\mu}\Phi\partial_{\nu}\Phi -
e^{-2\alpha \Phi}F_{\mu  \nu}F^{\mu \nu} ). \ee If $\alpha=0$ this
reduces to the EM system as the dilaton
decouples. Also note that if $\alpha=\sqrt{8/3}$, the theory
corresponds to the Kaluza-Klein (KK) theory one obtains by reducing
a $5+1$ dimensional pure gravity theory with one spacelike Killing
vector. It is convenient to define the parameter $N = 4/(\alpha^2
+4/3)$. The field equations are:
\bea
R_{\mu\nu} = 2\partial_{\mu}\Phi \partial_{\nu} \Phi + 2e^{-2\alpha \Phi} \left(F_{\mu \rho}F_{\nu}^{\phantom{\nu}\rho} - \frac{1}{6}g_{\mu\nu}
F_{\rho \sigma}F^{\rho\sigma} \right) \\ \nabla_{\mu}(
e^{-2\alpha\Phi} F^{\mu\nu})=0, \\
\nabla^2 \Phi +\frac{\alpha}{2}e^{-2\alpha\Phi}F^2 =0,
\eea
where the dilaton equation of motion
follows (if $\Phi$ is not a constant) from the other two. We have found
solutions to the theory, valid for all $\alpha$, which correspond to
static electrically charged dilatonic black rings: \bea \label{EMDring} ds^2 &=& -\frac{F(x)}{F(y)} \frac{dt^2}{
V_{\beta}(x,y)^{2N/3}} \\
\nonumber &+& \frac{R^2 V_{\beta}(x,y)^{N/3}}{(x-y)^2}
\left[ -F(x) \left( (1-y^2)d\psi^2+ \frac{F(y)}{1-y^2}dy^2 \right)+
F(y)^2 \left( \frac{dx^2}{1-x^2} + \frac{1-x^2}{F(x)}d\phi^2 \right)
\right] \\ A
&=& \frac{\sqrt{N}}{2} k\frac{(F(y)-F(x))}{F(y)-k^2F(x)} dt \\
e^{-\Phi} &=& V_{\beta}(x,y)^{N\alpha/4} \eea where $F(\xi) =
1-\lambda \xi$ and $k =\tanh \beta$ and $V_{\beta}(x,y)=
\cosh^2\beta -\sinh^2\beta \frac{F(x)}{F(y)}$. As in the case of the
neutral black ring the coordinate ranges are: $-\infty <t<\infty$,
$-1 \leq x \leq 1$, $- \infty < y \leq -1$ and $1/\lambda <y
<\infty$  and $\psi$ and $\phi$ are angles with periods which we
will determine. For the ring, the parameters take the ranges $0 <
\lambda < 1$ and $R>0$. Note that for $\alpha=0$ this is also a
solution of $D=5$ minimal supergravity since $F \wedge
F=0$. We should also emphasise that for $N=1,2,3$ our solutions
can be obtained as static limits of the rotating solutions of~\cite{ELVANGEMPARAN}
\footnote{The $N=3$ rotating charged solution is pathological due to the presence of Dirac-Misner strings.}.\\ \\

%\subsection{Physical properties}
It should be clear from the metric~(\ref{EMDring}) that we are
justified to interpret the solutions as static black rings. We now confirm this identification and proceed to compute
various physical quantities of our solution, such as the mass,
charge, surface gravity and area of the horizon. The physical
properties naturally share many of the same features as the
neutral static ring of~\cite{GEN}.
\par \noindent
First of all, it is clear that there is a horizon at $y = -\infty$. In the near-horizon limit, the metric
becomes (restricting to the $t-y$ plane)
\begin{equation}
ds^2_{ty} \sim F(x)(\cosh{\beta})^{2N/3} \left( -\frac{
  Y^2}{4R^2\lambda^2}d \tilde{t}^2 + dY^2 \right )
\end{equation} where $\tilde{t} \equiv ({\cosh{\beta}})^{-N}t$ and $y
  =-\frac{4R^2\lambda}{ Y^2}$. This is obviously conformal to Rindler spacetime (note $F(x) > 0$). Transforming to flat
  coordinates in the obvious way, we set $T =
  Y\sinh{\frac{1}{2R\lambda}}\tilde{t}$ and $X =
  Y\cosh{\frac{1}{2R\lambda}}\tilde{t}$, we see that the near horizon
  metric is non-singular:
\begin{equation}\label{horizon}
ds^{2}_{NH} \sim {(\cosh{\beta})}^{2N/3}\left( F(x) \left ( -dT^2 + dY^2 +
R^2 d \psi^2 \right ) + R^2\lambda^2 \left (
  \frac{dx^2}{1-x^2} + \frac{1-x^2}{F(x)}d\phi^2 \right) \right ).
\end{equation} On constant-time slices the
topology of the horizon is  $S^1 \times S^2$, parameterised by
$\psi$ and $(x,\phi)$ respectively, thus proving it is a black
ring\footnote{For
  $\lambda=1$ this analysis does not work and instead we recover the
  static, Einstein-Maxwell-dilaton black holes presented
  in~\cite{GIBBONSMAEDA}. One may see this explicitly by the
  coordinate change $r^2  =  \frac{4}{A^2}\frac{(1-y)}{(x-y)}$ and
$\cos^{2}{\theta}  =  \frac{(1-y)(1+x)}{2(x-y)}$.}. The situation is
completely analogous to the neutral static ring; as in that
particular case, we must demand that the period of $\psi$ satisfies
$\Delta \psi = 2\pi\sqrt{1 + \lambda}$ in order to avoid a conical
singularity at $y = -1$. The coordinate $x$ represents the polar
angle on $S^2$ and we are faced with either a conical singularity at
either the `north pole' at $x=-1$ or the `south pole' at $x=1$. The
apparent unavoidable existence of a conical singularity plagues all
known asymptotically flat static black rings; the problem is solved
in the rotating case, where the centripetal force manages to balance
the massive ring's self-attraction. It was hoped that the existence
of charge might create self-repulsion necessary to balance the
gravitational force, even in the absence of rotation. However, it is
clear this happy circumstance will not manifest itself here.
\par
\noindent
If we demand regularity at $x=-1$, then $\Delta \phi =
2\pi\sqrt{1+\lambda}$ and as we shall show, the solutions~(\ref{EMDring})
are asymptotically flat. The ring is said to be sitting on the rim of
disk shaped membrane with negative deficit angle. On the other hand,
to enforce regularity at $x=1$, we take $\Delta \phi =
2\pi\sqrt{1-\lambda}$, and the ring is said to be sitting on the rim of a
membrane that extends to infinity.
Computing the area of the horizon we find
\begin{equation}
A_{H\pm} = 8\pi^2R^3\cosh^{N}{\beta}\;\lambda^2\sqrt{(1+\lambda)(1
    \pm \lambda)}
\end{equation} where $\pm$ corresponds to taking the conical singularity at
    $x=\pm 1$.
\par
\noindent
Before considering the asymptotics of our solutions, we note that the near-horizon metric~(\ref{horizon}) is regular and we are entitled to
continue past the horizon $Y = 0$ and into the region $y \leq
\infty$. This is the `interior' part of the black ring solution, and we reach a true curvature singularity at $y = \frac{1}{\lambda}$, as
can be verified by computing the Kretchmann scalar explicitly.
\par
\noindent
As in~\cite{GEN}, one may verify that the metric~(\ref{EMDring}) is
asymptotically flat using the change of coordinates $\xi=
\frac{R(1+\lambda)\sqrt{y^2-1}}{(x-y)}$, $\eta =
\frac{R(1+\lambda)\sqrt{1-x^2}}{(x-y)}$, $\tilde{\psi}=\psi/ \sqrt{1+\lambda}$
and $\tilde{\phi}=\phi/ \sqrt{1+\lambda}$. Defining $\rho = \sqrt{\xi^2+\eta^2}$, we
see that as $\rho \to \infty$,
\bea
ds^2 \sim -dt^2+d\xi^2+d\eta^2 +\xi^2d\tilde{\psi}^2
+\eta^2d\tilde{\phi}^2
\eea
thus proving the assertion. Note that if the conical singularity lies
at $x=-1$ then the asymptotic metric is a `deficit membrane'.
At this stage we should remark that our solutions are consistent with
the uniqueness theorem of~\cite{uniqueness} which states that
asymptotically flat, regular, static dilaton black holes are
spherically symmetric. This because the assumption of regularity fails
due to the presence of a conical singularity.

\noindent In order to compute the mass of the
spacetime we will need the next to leading order term of
$g_{tt}$~\cite{MYERSPER}. Since
\be
g_{tt} \sim -\left( 1-2\frac{R^2\lambda(1+\lambda)}{
  \rho^2} \left(1+\frac{2N}{3}\sinh^2\beta\right)  + \cdots
\right), \ee the mass is therefore \be M_{\pm} = \frac{3\pi
R^2}{4}\lambda\sqrt{(1+\lambda)(1 \pm \lambda)} \left(
1+\frac{2N}{3}\sinh^2\beta \right). \ee One can extract the charge
from the asymptotic form of the potential $A$ which is, \be A \sim
\frac{\sqrt{N} k}{1-k^2} \frac{R^2\lambda(1+\lambda)}{\rho^2} dt \ee
and we define the charge by
$Q=\frac{1}{8\pi}\int_{\infty}dS_{\mu\nu}F^{\mu\nu}$ as
in~\cite{GMT}. Therefore, \be Q_{\pm}= \frac{\pi\sqrt{N} k}{1-k^2}
R^2\lambda\sqrt{(1+\lambda)(1\pm\lambda)}. \ee In the flat space
limit $\lambda = 0$ both the mass and charge of the ring  vanish.
This should be compared to the case of the standard
Reissner-Nordstrom solution in five dimensions. In that case
vanishing mass leads to a naked singularity for non-zero charge.
Note that in the EM theory ($N=3$) we have \be M_{\pm} =
\frac{\sqrt{3}}{4} \frac{1+k^2}{k}Q_{\pm}\geq
\frac{\sqrt{3}}{2}|Q_{\pm}| \ee with equality if and only if
$k^2=1$. This is consistent with the BPS bound derived
in~\cite{Gibbons}. Thus we expect a supersymmetric solution for this
value of $k^2$; however this is a delicate limit of the ring (since
$|\beta| \to \infty$) and one finds that one needs to send $\lambda
\to 0$ with $\lambda/(1-k^2)$ fixed~\cite{IDA}. After taking this
limit one is left with a spacetime with a naked singularity . Thus
there does not appear to be
an extremal ring in this class of solutions.\\
\par
\noindent
Given the existence of a horizon, one can compute the surface
gravity. We find that this is consistent with the temperature one gets
by Euclideanising the solution in the standard way, which is
\begin{equation}
T = \frac{1}{4\pi R\lambda\cosh^{N}{\beta}}.
\end{equation}
One can then easily find a Smarr relation
\begin{equation}
M = \frac{3}{8}TA_{H} +  Q \phi_H
\end{equation} where $\phi_H$ is the potential evaluated on the
horizon. We must note that these static black rings, which include the
neutral case of~\cite{GEN}, naively do not appear to satisfy the first
law of black hole mechanics\footnote{We thank H. S.
Reall for a discussion on this point.}. Presumably this failure is
related to non-zero contributions to the path integral from the
conical singularities, as of course the curvature tensors will include
a delta function supported on the deficit membrane. This would be an interesting puzzle to resolve
and would probably require techniques along the lines
of~\cite{BTZ}. \\ \\

%\section{A Harrison transformation}
The static solutions we have presented suffer from conical
singularities. Such solutions are presumably unstable. It was
remarked earlier that the presence of electric charge was not
sufficient to provide a repulsive force to balance the ring in the
way the centripetal acceleration caused by rotation does. However,
one might hope the presence of background electric fields may manage
to ``hold up'' the charged static ring. This is indeed the case. In
this section we apply a solution generating technique in order to
remove the conical singularity present in the black ring. The new
solution obtained will have the interpretation of a charged black
ring immersed in an electric field. A similar procedure was
performed in~\cite{emparan}, then later in~\cite{ortaggio} to the
dipole black rings in the pure EM case, although the resulting
background field was magnetic in both cases. The purpose of
performing the transformation is to alter the part of the metric
that causes the conical singularities, namely the $g_{\phi \phi}$
component. A ``standard'' Harrison transformation will do this (see
e.g.~\cite{dow}), however it requires that the $A$ potential is of
the form $A=A_{\phi}d\phi$. In contrast we have $A=A_t dt$. To
circumvent this problem we will consider the magnetic dual of our
solutions and then apply the formalism developed in~\cite{GALTSOV}.
To dualise the  solution one sets $H= e^{-2\alpha \Phi}*F$ and
$\tilde{\Phi}=-\Phi$ with the metric unchanged. Then one is dealing
with the theory: \be \label{dual} S = \frac{1}{16 \pi }\int d^5x
\sqrt{-g} \,\left(R -2(\partial \tilde{\Phi})^2-
\frac{1}{3}e^{-2\alpha \tilde{\Phi}}H^2 \right). \ee Note that the
Maxwell equations in one theory correspond to the Bianchi identity
in the other, thus $H=dB$. Our solution can be easily mapped to
solutions in the dual theory, and correspond to static black rings
sourced by a purely \emph{magnetic} potential given by \be B=
\frac{\sqrt{N} R^2\lambda k}{2(1-k^2)} \frac{(1-xy)}{(x-y)}d\psi
\wedge d\phi. \ee These rings may be contrasted with the rotating
black rings found in~\cite{MAGRING} which were charged electrically
by the $B$ field. The generating technique in~\cite{GALTSOV} is
suited to dealing with potentials with only one non-zero component
as is the case here. The metric will be written as \be ds^2=
\gamma_{\mu \nu}dy^{\mu}dy^{\nu} + (\sqrt{\gamma})^{-2} h_{\alpha
\beta}dx^{\alpha}dx^{ \beta} \ee where $y^{\mu} =  (\phi,\psi)$ and
$x^{\alpha} = ( t,x,y )$. The $B$ field will be denoted by $B_{\psi
\phi}= \frac{\sqrt{N}}{2} \Psi$. Also it is convenient to define
$\tilde{\gamma}_{\mu\nu} = \gamma_{\mu\nu}/ \sqrt{\gamma}$, $\chi =
2\alpha\tilde{\Phi} +2\log \sqrt{\gamma}$ and $\xi = 4\tilde{\Phi}
-3\alpha\log\sqrt{\gamma}$. In these variables one can show that the
action (\ref{dual}) for metrics possessing {\it two} commuting
Killing vectors (spacelike) reduces to a sigma model on
$[SL(2,\mathbb{R})]/SO(2)]^2 \times \mathbb{R}$. One can introduce
  ``Ernst'' potentials $\Psi$ and $\mathcal{E} = -e^{\chi}-
  \Psi^2$. There is a  ``Harrison type''
transformation that acts on $( \mathcal{E}, \Psi)$ as:
\bea
\mathcal{E}' = \frac{\mathcal{E}}{1-2c\Psi-c^2\mathcal{E}},
\qquad\Psi'= \frac{\Psi +c\mathcal{E}}{1-2c\Psi-c^2\mathcal{E}},
\eea
which is a symmetry of the sigma model.
It is this transformation which we will use. Note in particular that
$\xi$ and $\tilde{\gamma}_{\mu\nu}$ do not change under this
transformation. From this one may deduce how $\gamma_{\mu\nu}$
transforms:
\be
\gamma'_{\mu\nu} = \frac{\gamma_{\mu\nu}}{H(x,y)^{N/3}},
\ee
and how the dilaton transforms:
\be
e^{\tilde{\Phi}'} = \frac{e^{\tilde{\Phi}}}{H(x,y)^{N\alpha/4}},
\ee
where we have defined $H(x,y)=[(1-c\Psi)^2
  +c^2e^{\chi}]$.
For our solution
one may readily calculate:
\be
e^{\chi} = R^4\frac{F(y)^2(1-x^2)(y^2-1)}{(x-y)^4} V_{\beta}(x,y)^2.
\ee
The solution one obtains upon application of
the Harrison transformation is the following:
\bea
\label{melring}
ds^2 &=& -\frac{F(x)}{F(y)} \left(\frac{H(x,y)}{V_{\beta}(x,y)}\right)^{2N/3}dt^2 \\
\nonumber &+& \frac{R^2}{(x-y)^2}\left(\frac{ V_{\beta}(x,y)}{H(x,y)} \right)^{N/3}
\left[ -F(x) \left( (1-y^2)d\psi^2+
  \frac{H(x,y)^{N}F(y)}{1-y^2}dy^2 \right) \right.\\ && \qquad \qquad
  \qquad \qquad \qquad \qquad + \nonumber \left.
F(y)^2 \left( \frac{H(x,y)^{N}dx^2}{1-x^2} + \frac{1-x^2}{F(x)}d\phi^2 \right)
\right] \\
B_{\psi\phi}&=& \frac{\sqrt{N} R^2\lambda k}{2(1-k^2)H(x,y)}
\frac{(1-xy)}{(x-y)} \left( 1-c\frac{ R^2\lambda k}{(1-k^2)}
\frac{(1-xy)}{(x-y)} \right) - \frac{\sqrt{N}}{2H(x,y)}ce^{\chi} \\
e^{\tilde{\Phi}}&=& \left( \frac{V_{\beta}(x,y)}{H(x,y)} \right)^{N\alpha/4}
\eea
Let us explore the properties of this solution now. First we will
consider the metric as $y \to -\infty$. Observe that in this limit
\be
H(x,y) \to \left[
  \left(1-cR^2\lambda \sinh\beta\cosh\beta \; x \right)^2 +c^2\cosh^4\beta
    R^4\lambda^2(1-x^2) \right] \equiv H(x). \ee
The metric in this limit looks rather similar to our original solution,
and thus we deduce (for $0<\lambda<1$) that this new metric has a non-singular horizon of
topology $S^2 \times S^1$,
since $H(x) > 0$ for $ -1 \leq x \leq 1$. One may check that the
conical singularities, in the $x\phi$ part of the metric, at $x=\mp 1
$ are removed by identifying $\phi$ with period
\be
\Delta \phi = 2\pi\sqrt{1\pm \lambda}\left( 1\pm \frac{cR^2\lambda
  k}{(1-k^2)} \right)^{N}
\ee
This allows one to solve for the parameter $c$ which we find to be:
\be
c= \frac{(1-k^2)}{R^2\lambda k} \left(\frac{1- \left(
  \frac{1+\lambda}{1-\lambda} \right)^{1/2N}}{1+ \left(
  \frac{1+\lambda}{1-\lambda} \right)^{1/2N}}\right) \equiv
\frac{(1-k^2)}{R^2\lambda k} \varpi \ee which can be interpreted as
the critical value of the strength of the form field required to
keep the charged ring balanced. Note that $c$ cannot be defined in
the case of zero charge, i.e. $k=0$. This lends support to the
physical interpretation that only a charged ring could be held in
equilibrium by a field. We have defined the quantity $\varpi$ for
convenience - note in particular that since $0<\lambda <1$ it
follows that $-1<\varpi < 0$. Finally, the conical singularity at $y
= -1$ occurring in the $g_{y\psi}$ part of the metric can be removed
by the identification
\begin{equation}
\Delta \psi = 2\pi\sqrt{1 + \lambda} \left( 1 - \varpi \right)^{N}.
\end{equation}
It is now clear that the metric (\ref{melring}) represents a black ring. However it is
not asymptotically flat. Thus we interpret this as the ring
``immersed'' in some non-trivial background. To determine this
background we simply set $\lambda=0$.  We may consider this background
as the effect of the Harrison transformation on flat
spacetime. Note that there are no conical singularities and the
parameter $c$ is again a free parameter. Defining radial and polar
coordinates as
\begin{equation}
\xi^2=\rho^2\sin^2{\theta} = R^2\frac{(y^2 -1)}{(x-y)^2}, \phantom{===}
\eta^2=\rho^2\cos^2{\theta} = R^2\frac{(1-x^2)}{(x-y)^2},
\end{equation} the background metric takes the form of a dilaton fluxbrane
\begin{equation}
\label{fluxbrane}
ds^ 2 = H^{2N/3}(-dt^2 + d\rho^2 + \rho^2 d\theta^2) +
H^{-N/3}\rho^2(\sin^2{\theta} d\psi^2 + \cos^2{\theta} d\phi^2)
\end{equation} with $H(\rho,\theta) = (1 + c^2 \rho^4\sin^2{\theta}
\cos^2{\theta})$. The two-form potential is
\begin{equation}
B_{\psi \phi} =
-\frac{\sqrt{N}c \, \xi^2\eta^2}{2(1 + c^2\xi^2\eta^2)}.
\end{equation}  At infinity, $B$ clearly goes to a constant
that cannot be removed by a gauge transformation. The dilaton reads
\begin{equation}
e^{\tilde{\Phi}} = H(\rho,\theta)^{-N\alpha/4}.
\end{equation}
One may now readily calculate the Maxwell potential $A$ in the dual
(EMD) theory to find \be A= \frac{\sqrt{N}c}{2}(\xi^2-\eta^2)dt. \ee
This corresponds to the electric field which distorts the background
(\ref{fluxbrane}). One may go on to calculate the physical
quantities for the ring immersed in the electric field. The only
difference one finds, as compared to the asymptotically flat case,
arises from the ranges of the $\psi, \phi$ angles now being
different. Also, of course, to compute the mass and charge of the
ring, one needs to ``subtract'' the effect of the background values
of the metric and potential.

\noindent Finally note that for $\lambda=1$ the above analysis would generate a
static dilatonic charged black hole (which includes the
Reissner-Nordstrom solution) in the electric background above. \\ \\

%\section{Discussion}
We conclude with some comments on the problem of adding rotation to
our solutions.
One of the original motivations for considering this problem was to
find a Myers-Perry black hole electrically charged under the Maxwell
field in five dimensions. Our hope was to first find a charged,
rotating ring and then take an appropriate limit to leave a black
hole, as can be done in the uncharged case. It would thus be interesting
to find rotating generalisations of our solutions.
Of course we can easily give such a solution in the case of
Kaluza-Klein coupling ($\alpha = \sqrt{8/3}$). It may be
deduced from the more generalised solutions found in~\cite{ELVANGEMPARAN} where
it was considered in the dualised theory. Here, however, we consider
it in the context of rotating black rings of EMD theory. As is
well known, the procedure consists of lifting the vacuum solution to one higher dimension by adding a
flat direction (with coordinate $z$ say), then applying a boost $(t,z)
\to ( t\cosh\beta + z\sinh \beta , t\sinh\beta + z\cosh\beta )$.
Finally one drops back down a dimension reducing along the $z$
direction. This will generate a non-zero dilaton and gauge
field. Applying this to the 5D vacuum rotating black ring
of~\cite{ROT} we find
\bea
ds^2 &=& -\frac{F(x)}{F(y)}\frac{(dt+R\sqrt{\lambda\nu}(1+y)d\psi)^2 }{
V_{\beta}(x,y)^{2/3}} \nonumber\\&& \nonumber+\frac{R^2}{(x-y)^2}V_{\beta}(x,y)^{1/3}\\&& \times \left[ -F(x) \left( G(y)d\psi^2+ \frac{F(y)}{G(y)}dy^2 \right)+
F(y)^2 \left( \frac{dx^2}{G(x)} + \frac{G(x)}{F(x)}d\phi^2 \right)
\right], \\
 A &=& \frac{1}{2} k\frac{(F(y)-F(x))}{F(y)-k^2F(x)} dt
 -\frac{R\sqrt{\lambda\nu}k(1-k^2)(1+y)F(x)}{F(y)-k^2F(x)}d\psi, \\
e^{-\Phi}&=& V_{\beta}(x,y)^{-\frac{1}{4}\sqrt{8/3}}
\eea
where $G(\xi)=(1-\xi^2)(1-\nu(1-k^2)\xi)$, $k=\tanh \beta$; note that
we have redefined the parameter $\nu$ appearing in the vacuum black
ring to cast it into this form, such that $\nu_{here} =
  \nu_{vacuum}\cosh^2\beta$. This solution of course reduces to the
static black ring for $\alpha=\sqrt{8/3}$ we presented in the previous
section when $\nu=0$.
 While an exact rotating solution for arbitrary dilaton coupling
 remains to be found, one can try to follow the method of Horne
and Horowitz~\cite{HOHO}. In their attempt to find a rotating dilaton
black hole in four dimensions, they observed that one could solve the
Einstein-Maxwell-dilaton equations to linear order in the rotation
parameter. They actually already knew the exact solution to the problem for two
values of the dilaton coupling - the Kaluza-Klein case and the
Kerr-Newmann black hole. Note this is in contrast to the present
case.
It is clear from the above metric that, to linear order in the
rotation parameter, only the $g_{t\psi}$ component of the metric is
affected. A natural ansatz for the perturbation about the static solution would therefore be
$g_{t \psi} = \omega f(x,y)(1+y)$ and $A_{\psi} = \omega
g(x,y)(1+y)$ where $\omega$ denotes the rotation parameter and we work
to $O(\omega)$. We found that for
\begin{eqnarray}
f(x,y) & = &\frac{F(x)}{F(y)V_{\beta}(x,y)^{2N/3}} \\
g(x,y) & = & \frac{\sqrt{N}}{2}\frac{k(k^2-1)F(x)}{F(y)-k^2F(x)}
\end{eqnarray} the Einstein equations are satisfied; however, the
  Maxwell equations are only satisfied for
  $\alpha=\sqrt{8/3}$. Further, in the EM case ($\alpha=0$), one may
  also include a Chern-Simons term to (\ref{action}). However one
  finds that this does not work either. This is because only the $\psi$
  component of the linearised Maxwell equation is non-zero, whereas
  only the $\phi$ component of the linearised Chern-Simons term
  $*(F\wedge F)$ is non-zero.   Thus it seems probable that an ansatz of the
  above form will not work, and one will probably need to include rotation
  in the $\phi$ direction too. This would be consistent with a
  non-extremal version of the supersymmetric black ring~\cite{SUSYRING}.
  In any case all this seems to suggest that a Chern-Simons term is required
  in the EMD theory in order to obtain rotating black rings.\\ \\

\noindent {\bf Note added:}
We have just become aware that a non-extremal charged rotating black
ring in the EM theory with Chern-Simons term has been found
in~\cite{eef}, as anticipated above. This solution possesses one
independent rotation parameter. However the most general rotating
solution remains to be found. \\ \\

\noindent {\bf Acknowledgment:} The authors would like to thank Gary
Gibbons, Malcolm Perry and Bella Schelpe. The authors would also
like to thank Roberto Emparan for a number of useful comments. HKK
would like to thank St. John's College, Cambridge, for financial
support.

%%%%%%%%%%%%%%%%%%%%%%%%%%%%%%%%%%%%%%%%%%%%%%%%%%%%%%%%%%%%%%%%%%%%%%%%
%                       REFERENCES                                     %
%%%%%%%%%%%%%%%%%%%%%%%%%%%%%%%%%%%%%%%%%%%%%%%%%%%%%%%%%%%%%%%%%%%%%%%%
%\newpage

\renewcommand{\thefootnote}{\arabic{footnote}}
\setcounter{footnote}{0}

%\tableofcontents
%%%%%%%%%%%%%%%%%%%%%%%%%%%%%%%%%%%%%%%%%%%%%%%%%%%%%%%%%%%%%%%%%%%%%%
%\section{Introduction}
%%%%%%%%%%%%%%%%%%%%%%%%%%%%%%%%%%%%%%%%%%%%%%%%%%%%%%%%%%%%%%%%%%%%%%

\end{document}